\documentclass{aa}  

\usepackage{graphicx}
\usepackage{txfonts}
\def\fmax{f_{\rm max}}
\def\rhom{\overline\rho}

\def\msun{M_\odot}

\def\mmax{M_{\rm max}^{\rm stat}}
\usepackage{colortbl}

\bibpunct{(}{)}{;}{a}{}{,} 
\begin{document}

   \title{Fundamental physics and the absence of sub-millisecond pulsars}
\authorrunning{B.~Haskell et al.}
   \subtitle{}

   \author{B.~Haskell \inst{1} \fnmsep\thanks{email: bhaskell@camk.edu.pl}, J.~ L.~Zdunik\inst{1}, M.~Fortin\inst{1}, M.~Bejger\inst{1,2}, R.~Wijnands\inst{3} \and A.~Patruno\inst{4}}

   \institute{Nicolaus Copernicus Astronomical Center, Polish Academy of Sciences,
Bartycka 18, 00-716 Warszawa, Poland\\
\and APC, AstroParticule et Cosmologie, Universit\' e Paris Diderot, CNRS/IN2P3, CEA/Irfu, Observatoire de Paris, Sorbonne Paris Cit\'e, F-75205 Paris Cedex 13, France\\
      \and Anton Pannekoek Institute for Astronomy, University of Amsterdam, Science Park 904, 1098 XH, Amsterdam, the Netherlands\\
         \and Leiden Observatory, Leiden University, P.O Box 9513, 2300 RA, Leiden, the Netherlands}

   \date{Received ??; accepted ??}

 
  \abstract
   {Rapidly rotating neutron stars are an ideal laboratory to test models of matter at high densities. In particular the maximum rotation frequency of a neutron star is equation of state dependent, and can be used to test models of the interior. However observations of the spin distribution of rapidly rotating neutron stars show evidence for a lack of stars spinning at frequencies larger than $f\approx 700$ Hz, well below the predictions of theoretical equations of state. This has generally been taken as evidence of an additional spin-down torque operating in these systems and it has been suggested that gravitational wave torques may be operating and be linked to a potentially observable signal.}
   {In this paper we aim to determine whether additional spin-down torques (possibly due to gravitational wave emission) are necessary, or whether the observed limit of $f\approx 700$ Hz could correspond to the Keplerian (mass-shedding) break-up frequency for the observed systems and is simply a consequence of the, currently unknown, state of matter at high densities.}
   {Given our ignorance with regard to the true equation of state of matter above nuclear saturation densities, we make minimal physical assumption and only demand causality, i.e. that the speed of sound in the interior of the neutron star should be less or equal to the speed of light $c$. We then connect our causally-limited equation of state to a realistic microphysical crustal equation of state for densities below nuclear saturation density. This produces a limiting model that will give the lowest possible maximum frequency, which we compare to observational constraints on neutron star masses and frequencies. We also compare our findings with the constraints on the tidal deformability obtained in the observations of the GW170817 event.}
   {We find that the lack of pulsars spinning faster than $f\approx 700$ Hz is not compatible with our causal limited `minimal' equation of state, for which the breakup frequency cannot be lower than $f_{\rm max}\approx 1200$ Hz. A low frequency cutoff, around $f\approx 800$ Hz could only be possible if we assume that these systems do not contain neutron stars with masses above $M\approx 2 M_\odot$. This would have to be due either to selection effects, or possibly to a phase transition in the interior of the neutron star, that leads to softening at high densities and a collapse either to a black hole or a hybrid star above $M\approx 2 M_\odot$ . Such a scenario would, however, require a somewhat unrealistically stiff equation of state for hadronic matter.}
  {}


   \maketitle
%

\section{Introduction}

Neutron Stars (NSs) are extraordinary cosmic laboratories, as they allow to probe aspects of all fundamental interactions at extremes of density and gravity impossible to reproduce in terrestrial experiments.
In fact, while current methods allow to study the physics of crustal layers below nuclear saturation density $\rho_0=2.7 \times 10^{14}$ g cm$^{-3}$, the density in the core of a NS can exceed $\rho_0$ by an order of magnitude, entering a highly uncertain regime well beyond the possibilities of rigorous ab-initio theoretical modelling (for a textbook introduction, see \citealt{Haensel2007}).

Constraints on theory have come from the observation of NSs with masses close to $2 M_\odot$: PSR J1614-2230~\citep{Demorest2010,Fonseca2016,Arzoumanian2018}
 and PSR J0348+0432~\citep{Antoniadis2013}, as well as recent indications that PSR J2215+5135 may have a mass of $M\approx 2.3 M_\odot$ \citep{Linares}. These systems allow us to constrain several models for the Equation Of State (EOS) of dense matter, by imposing that matter in the core must be stiff enough to allow for a maximum mass $M_{\rm max}>2 M_\odot$ (see e.g. \citealt{Fortin2016}). More stringent constraints are expected in the near future, when the NICER \citep{Arzoumanian2014}, and further in the future ATHENA \citep{Motch2013}, missions' data will allow to accurately determine pulse profiles and thus not only measure the mass, but also constrain the radius of several NSs.

While simultaneous measurements of mass and radius would allow for strong constraints on the EOS, it is important to keep in mind that many NSs rotate rapidly, with frequencies $f$ of up to $f\approx 700$ Hz \citep{Papitto14, Hessels, Haensel2016, Patruno2017, Bassa}. Rapid rotation leads to increased centrifugal support, and allows for the existence of supramassive NSs (objects with a mass higher than the maximum mass of the static configuration). This not only alters the mass-radius curve for a given EOS, and must be kept in mind when analysing observational data, such as will be provided by NICER \citep{Magda}, but may lead to a spin-down induced collapse which has been suggested as a possible engine for energetic phenomena such as fast radio bursts \citep{FR2014, RL2014}, or the short-lived X-ray afterglows observed in some short gamma ray bursts \citep{DL1, DL2, ZM2001, GF2006, Lasky14}.

However NS rotation also opens up the possibility of further, independent constraints on the EOS. First of all, accurate timing of pulsars in compact binaries over long periods can constrain the spin-orbit coupling and allow for measurements of the moment of inertia of the NS (see e.g. \citealt{BejgerBH2005,KramerWex,Watts2015,PJ17}). Another potential probe of NS structure and EOS is the determination of the mass-shedding limit, i.e. the maximum rotation rate the star can sustain before mass is stripped from the outer layers. Measurements of spins of millisecond radio pulsars and accreting millisecond X-ray pulsars are then a valuable tool to place lower limits on the maximum rotation rate of a NS.

While theoretical models based on state of the art modelling of the EOS for hadronic matter all predict maximum frequencies well above 1 kHz (see e.g. \citealt{Haensel2009}), observations in radio and in X-rays, however, have not led to the discovery of any NS rotating with sub-millisecond periods, with the fastest known pulsar, PSR J1748-2446ad, rotating at 716 Hz \citep{Hessels2006}. In particular a recent analysis of the spin distribution of accreting NSs in Low Mass X-ray Binaries (LMXBs) by \citet{Patruno2017} revealed also that the distribution is well described by two populations, a slower one that peaks around 300 Hz and a much narrower (standard deviation $\sigma\approx 30$ Hz) one of fast systems above 540 Hz, with an average frequency of 575 Hz.
Standard accretion models, even accounting for the transient nature of the spin-up over repeated accretion episodes, struggle to explain the lack of fast pulsars \citep{Patruno2017, Dangelo2017, B2017} and it has thus been suggested that an additional spin-down torque, that grows rapidly for high-frequencies, is required to explain it (although see \citealt{ABCPHD} for a discussion of how accretion torques alone can explain the data). Gravitational Waves (GWs) provide a natural mechanism, and it has been suggested that the lack of sub-millisecond pulsars is an indication of quadrupolar `mountains' in the crust or of unstable modes in the core, radiating gravitationally at a level sufficient to balance the spin-up torque due to accretion \citep{PP1978, Wagoner1984, Bildsten1998, Nils1998}. This scenario is, however, highly uncertain and estimates of GW emission from known systems, based on theoretical models, indicate a much lower level of emission in most cases \citep{Haskell15, Haskell17}, while current GW detectors, such as Advanced LIGO and Virgo, are not yet sensitive enough to detect the waves directly \citep{LSC2017a}.

In this paper we approach the problem from a different angle. Rather than assume a theoretical EOS, and assume an additional spin-down torque to stop the NS from spinning up to their breakup frequency, we remain agnostic with regards to the EOS and assume that the NSs have spun-up to their maximum frequency, the mass-shedding frequency. Our goal is thus to understand whether the lack of rapidly rotating pulsars (above $f\approx 700$ Hz) can be explained without additional spin-down torques, and determine what this would mean for the physics of the high density interior of the star.
To do this we follow the approach of \citet{Koranda1997} (see also \citealt{ILQ2018}) and first of all establish how low the break-up frequency can be by making only basic physical assumptions, i.e. imposing simply that the EOS remain causal in the core of star. Below the nuclear saturation density we adopt a realistic crust EOS, as will be described in the following sections. We analyse the impact on the stellar rotation rate, by also imposing that the maximum mass of a NS has to be at least the currently observed maximum of $2 M_\odot$ and that the mass-shedding frequency must exceed the currently observed maximum rotation rate of $f\approx 716$ Hz.

We then consider whether the lack of NS spinning faster than $f\approx 700$ Hz is consistent with causality and our understanding of low density physics. In general we find that it is inconsistent with our minimal physical assumptions and that additional physics is needed, either in the form of exterior spin-down torques (e.g. due to GW emission), or in the form of a phase transition in the interior of the star, that softens an otherwise stiff hadronic equation of state at high densities, and leads to collapse (either to a black hole or a hybrid star) if accretion pushes the stellar mass above $M\approx 2 M_\odot$ \citep{Bejger2017twins}.

\section{The maximum rotation frequency of neutron stars}

The maximum rotation rate of a NS is determined by the frequency $\fmax$ of a test particle co-rotating at the star's equator: this is the so-called mass-shedding frequency limit, above which matter is no longer bound and is ejected from the outer layers of the star. The frequency $\fmax$ at the equator depends on the gravitational mass of the star and is EOS dependent. The determination of $\fmax$ generally requires the calculation of rotating general relativistic equilibrium configurations. However \citet{Haensel2009} have shown that to a good degree of accuracy the mass-shedding frequency $\fmax$ can be determined by the EOS independent empirical formula:

\begin{equation}
 \fmax(M)=C~\left({M\over
M_\odot}\right)^{1/2}\left({R\over 10~{\rm
km}}\right)^{-3/2}~,
\label{eq:fmax}
\end{equation}
%

where $M$ is the gravitational mass of the \emph{rotating} star, $R$ is the radius of the \emph{non-rotating} star of mass
$M$, and $C$ is a prefactor, which is approximately EOS independent and only depends on whether we are dealing with a hadronic star (low density at the surface), or a self-bound (strange) star. For standard hadronic matter the coefficient is $C_{NS}=1.08~$kHz, while for self bound strange stars (with a hadronic outer crust) it is $C_{S} =1.15~$kHz, with an accuracy of a few percent as discussed below \citep{Haensel2009}.  The second value of $C$ is also appropriate for any EOS that is self-bound at high density (of the order of nuclear
matter density) and thus should be used also for the limiting EOS we construct in the following, the so-called `causal' EOS, that is maximally stiff and corresponds to matter in which the speed of sound is equal to the speed of light. The difference between the two values of $C$ is an example of the role of the outer layers of the NS (the crust), as their response to fast rotation in the equatorial plane is stronger than that of the NS core, thus making a quantitative difference in the case of strange stars with only an outer crust (for a detailed analysis see \citealt{Zdunik2001}, where the role of
the outer crust on rotating strange stars was discussed). The formula above can be used for masses up to $0.9M^{\rm stat}_{\rm max}$, with $M^{\rm stat}_{\rm max}$ the maximum mass of the non-rotating configuration. Note that, as will be discussed in the following, this precludes us from using it to estimate the absolute maximum mass-shedding frequency for a given EOS, as this will generally correspond to the maximum mass configuration. Nevertheless Eq. (\ref{eq:fmax}) is an important tool to estimate the mass-shedding frequency of the majority of the configurations, and the average density in the interior of the NS.

The relative errors when comparing results of Eq. (\ref{eq:fmax}) to numerical calculations are typically within 2\%, with largest deviations of $5\%$ at highest masses, although the precision of the formula worsens
slightly on the low-mass side, for masses below $0.5~M_\odot$. A detailed discussion of Eq.~(\ref{eq:fmax}) for different EOSs and the accuracy of these approximations is presented in \citet{Haensel2009}.
It can be inverted to obtain the mean density of the non-rotating star for which the mass-shedding frequency is $\fmax$ at given mass.
\begin{equation}
\rhom=1.76\rho_0 ~\frac{\fmax^2}{C^2}~, \label{meand}
\end{equation}
which decreases with decreasing maximum frequency. In Table~(\ref{density}) we show the mean densities and radii of maximally rotating configurations with a mass $M=2 M_\odot$. It is clear that low maximum frequencies, close to the currently observed maximum of 716 Hz, will lead to very low average densities, below the nuclear saturation density $\rho_0$. As we will see in the following, in this range of densities the EOS for the crust is constrained well enough to limit how stiff the EOS can be and the possibility of having low values for the breakup frequency.

\begin{table}
\caption{Mean density (in units of the saturation density $\rho_0=2.7 \times 10^{14}$ g cm$^{-3}$) and stellar radius (in km) of a NS with $M=2\msun$ for the two values of $C$ entering in Eq.\;(\ref{eq:fmax}) and for $\fmax$ equal to the largest observed spin frequency (716 Hz) and to 800 Hz.}
\label{density}   
\centering          
\begin{tabular}{l|ll|ll} 
\hline\hline     
$\fmax$ & \multicolumn{2}{l|}{$C=1.08$ kHz} & \multicolumn{2}{l}{$C=1.15$ kHz} \\ 
        & $\rhom$ &$R~(2\msun)$& $\rhom$ &$R~(2\msun)$\\
\hline                    
716&0.77 &16.56 &0.68 &17.28\\
800&0.97 &15.39 &0.85 &16.04\\
\hline                  
\end{tabular}
\end{table}

As can be seen from Eq.\;(\ref{eq:fmax}) the mass-shedding frequency increases with the mass, and is thus maximum for the maximum mass $M$ that can be obtained for a given EOS. This value depends on the choice of EOS, and is furthermore beyond the range of validity of Eq.\;(\ref{eq:fmax}), which can only be used reliably in the mass range $0.5 M_\odot <M < 0.9M_{\rm max}^{\rm stat}$. 
To obtain an absolute upper bound on $f_{\rm max}$ one can however fit maximum mass configurations of NSs rotating at the breakup frequency and obtain the following expression for the absolute maximum frequency $f_{\rm max} ^{\rm EOS}$:
\begin{equation}
f_{\rm max}^{\rm EOS}=C_{\rm max}~\left({\mmax\over
M_\odot}\right)^{1/2}\left({R^{\rm stat}_{M_{\rm max}}\over 10~{\rm
km}}\right)^{-3/2}~,\label{f-max-eos}
\end{equation}
where $C_{\rm max}=1.22$~kHz \citep{Haensel1995} and $R^{\rm stat}_{M_{\rm max}}$ is the radius of the non-rotating star at the maximum mass $\mmax$. $C_{\rm max}$ is independent of the EOS , while the maximum mass $M_{\rm max}$  (and corresponding radius) clearly depends on it. Note that close to the mass-shedding frequency the maximum mass of the rotating star $M_{\rm max}$ will generally be higher than the maximum mass of the non-rotating configuration $M^{\rm stat}_{\rm max}$ by up to 20\% \citep{Lasota1996}. To account for this and estimate the accuracy of the formula in (\ref{f-max-eos}) we have also calculated maximally rigidly rotating models for the EOSs described below with the use of the multi-domain spectral methods library {\tt LORENE}\footnote{\tt http://www.lorene.obspm.fr}  \citep{Gourgoulhon2016} and the {\tt nrotstar} code \citep{BonazzolaGSM1993, Gourgoulhon99}. { We also recall the result of \citet{BejgerHZ2007}, where it was demonstrated that with a great accuracy the mass-shedding frequency may be approximated by 
\begin{equation} 
f_{\rm max} = \frac{1}{2\pi}\sqrt{\frac{GM}{R^3_{\rm eq}}}, 
\label{eq:bhz2007} 
\end{equation} 
with gravitational mass $M$ and equatorial (circumferential) radius $R_{eq}$ of a star rotating at the mass-shedding frequency $f_{\rm max}$. This relation may be used to estimate the maximum radius that a star can support for a given $f_{\rm max}$ and $M$; for $f_{\rm max} = 716$ Hz one gets 
\begin{equation} 
R^{\rm max}_{\rm eq} = 20.94 \left(\frac{M}{1.4\,M_\odot}\right)^{1/3}\mathrm{km}. 
\label{eq:bhz2007_rmax}
\end{equation} 
}

\section{Causal Limit EOS}

Given our ignorance regarding the EOS of matter at supranuclear densities, we will remain agnostic regarding the true EOS and only demand causality throughout the star. As a limiting case we thus consider the maximally compact Causal Limit (CL) EOS:
\begin{equation}
P=(\rho-\rho_u)c^2,
\end{equation}
for which the speed of sound $c_s$ is such that 
\begin{equation}
c_s^2=\frac{dP}{d\rho}=c^2,
\end{equation}
where $P$ is the pressure and $\rho_u$ is the minimum mass-energy density to which the CL EOS extends, which in our case will be a free parameter. For self bound stars $\rho_u>0$ while for stars with a crust, $\rho_u$ is approximately the density at the crust-CL EOS boundary. 
For the CL EOS the maximum  mass $M_{\rm max}$ of a non-rotating self-bound star is \citep{Glendenningbook}:
\begin{equation}
\label{max-causal} 
M_{\rm max}=4.07\sqrt{\frac{2.7\cdot 10^{14} \mbox{~g~cm$^{-3}$}}{\rho_u}}~\msun~,
\end{equation}

with a radius
\begin{equation}
\label{max-causalrad}
R_{\rm Mmax}=17.1\sqrt{\frac{2.7\cdot 10^{14}\mbox{~g~cm$^{-3}$}}{\rho_u}}~{\rm km}~.
\end{equation}

Let us begin by assuming that the maximum spin frequency is the one of the fastest observed rotating NS $f_{\rm max}=716$ Hz. From Eqs.\;(\ref{meand}) and (\ref{f-max-eos}) we obtain a mean density of $\rhom=0.61\rho_0$, which from Eqs.\;(\ref{max-causal}) and (\ref{max-causalrad}) results in $\rho_u=0.43 \rho_0$. This density is well below the nuclear saturation density and would thus imply that regions of the star with $\rho<\rho_0$ are also described by the CL EOS. This however can be excluded from  our understanding of low density NS physics, as despite significant uncertainties regarding composition and transport properties, the stiffness of the crustal EOS is constrained well enough to discard the CL EOS as a viable option in this region \citep{Fortin2016,Haensel2017}. 


If we accept that our ignorance of the EOS begins at densities $\rho>\rho_0$, it is thus necessary to construct configurations with $\rho_u>\rho_0$, and supplement the CL EOS with a realistic crust for lower densities.

Before doing this we note that if in Eq. (\ref{max-causal}) we take $\rho_u=\rho_0$ the maximum mass is well above the observed maximum of $\approx 2 M_\odot$, and the radius of such a configuration is larger than the range currently suggested by X-ray observations, of between 10 and 14 km \citep{Ozel2016, Haensel2016}.

   \begin{figure}
   \centering
  \includegraphics[width=\hsize]{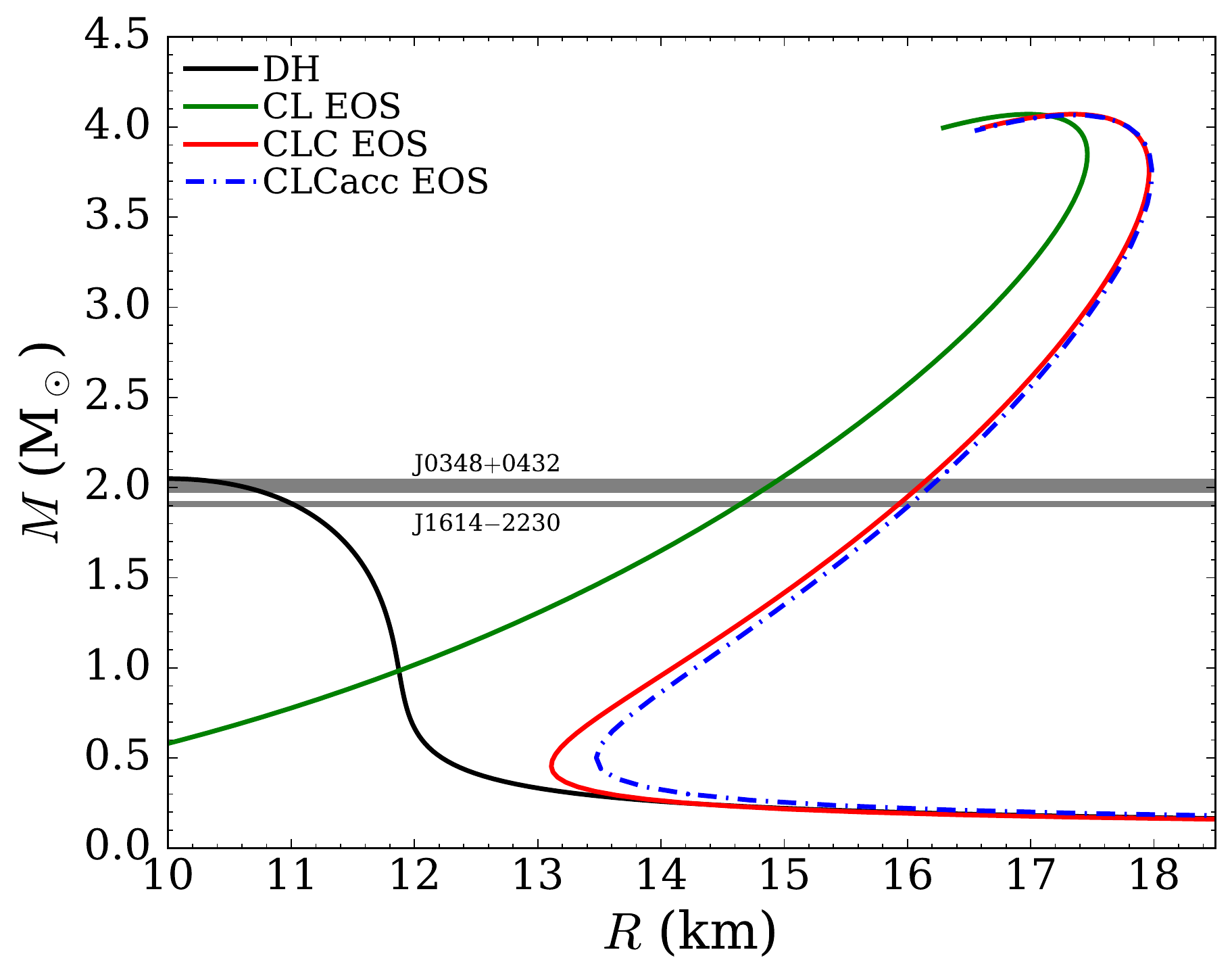}
      \caption{Mass-radius relations for non-rotating stars for the CL EOS with $\rho_u=\rho_0$, the CL EOS with { two crust models connected at $\rho_u=\rho_0$:  the catalyzed crust from \cite{DH} (CLC EOS) and the accreted crust from \cite{HZ08} (CLCacc EOS)}. For reference we plot the $M-R$ relation obtained for the DH EOS. The horizontal lines correspond to the largest observed maximum mass for PSR J1614-2230 and PSR J0348+0432.}
         \label{FigMR}
   \end{figure}

   \begin{figure}
   \centering
  \includegraphics[width=\hsize]{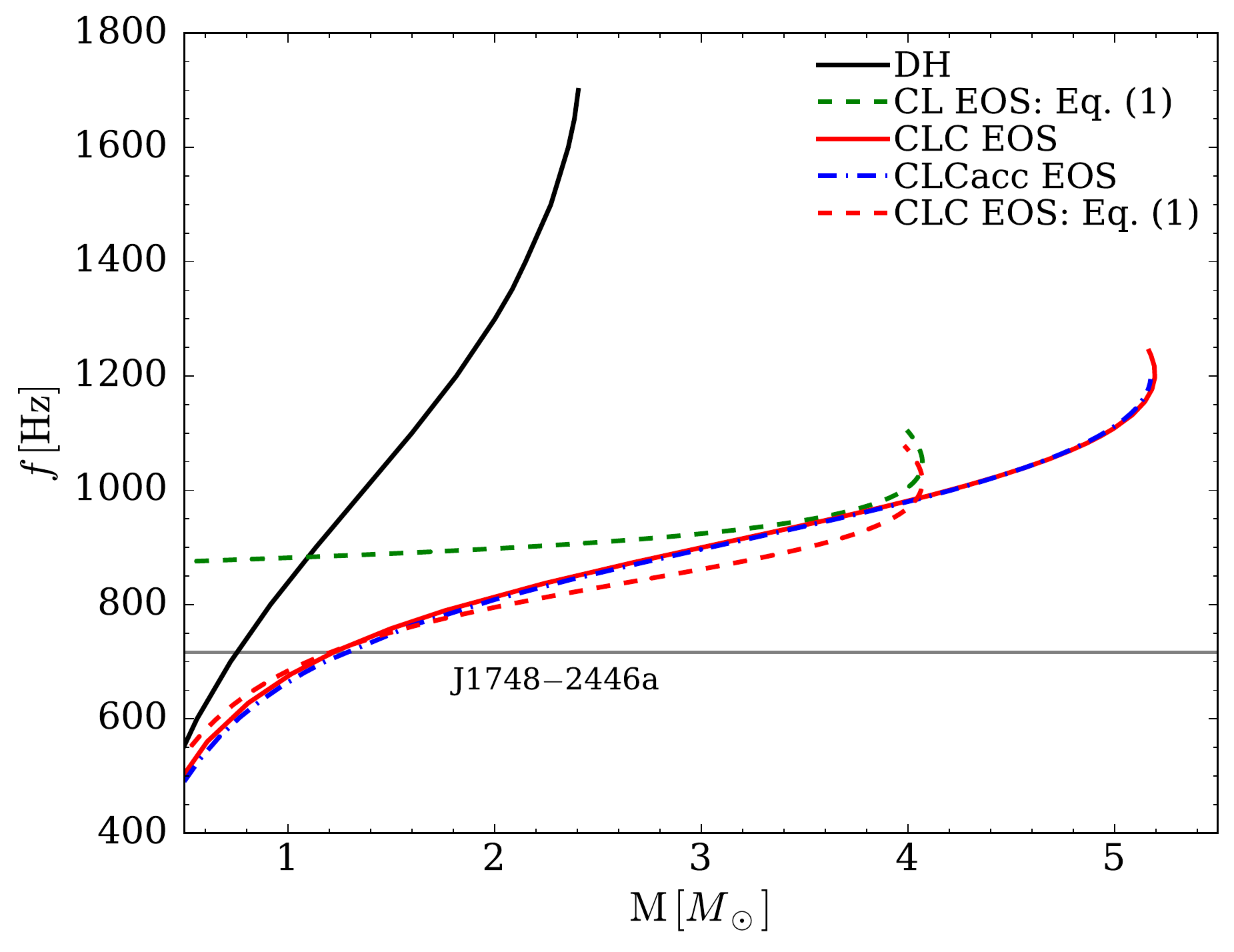}
      \caption{Mass-shedding frequency vs mass for the same EOSs as in Fig.\;\ref{FigMR}. For the { CLC EOS} we plot both the curves obtained with the approximate expression in Eq. (\ref{f-max-eos}) from a static configuration and exact numerical solutions for rotating configurations obtained with {\tt LORENE}. For the DH EOS, configurations also obtained with {\tt LORENE} are plotted. It can be seen that the maximum mass close to the mass-shedding frequency increases significantly due to the additional rotational support, but the error on the frequency remains small, around $10\%$. We see that for the most extreme configuration that is physically plausible, the CL EOS with crust, the mass-shedding frequency is around $1200$ Hz, well above the observed maximum of $f=716$ Hz. As this is the stiffest equation of state that can be built, any softening will lead to a higher mass-shedding frequency, as can be seen by comparing to the curve for the more realistic DH equation of state, which reaches mass shedding around $f\simeq 1730$ Hz.}
         \label{FigfM}
   \end{figure}

\section{Including the crust}

To obtain a robust limit of $f_{\rm max}$ we consider models in which our ignorance is restricted to densities $\rho>\rho_0$, i.e. to the core of the NS. For the core we use the CL limit EOS $P=(\rho-\rho_u)c^2$ described previously, with $\rho_u=\rho_0(1-P_0/\rho_0 c^2)$ and $P_0$ the pressure at the crust-core boundary \footnote{Strictly speaking our construction, that requires continuity of both $P$ and $\rho$, leads to $\rho_u$ being smaller than $\rho_0$ by approximately 1$\%$.}. For the crust we use a standard EOS from \cite{DH} (DH in the following) for matter below the nuclear matter density $\rho_0$ (corresponding to $n_0=0.16~{\rm fm^{-3}}$), and connect it to the core EOS assuming continuity of the chemical potential $\mu$ and pressure $P$ at the crust-core interface, following the approach presented in \cite{Zdunik2017}.

   \begin{figure}
   \centering
  \includegraphics[width=\hsize]{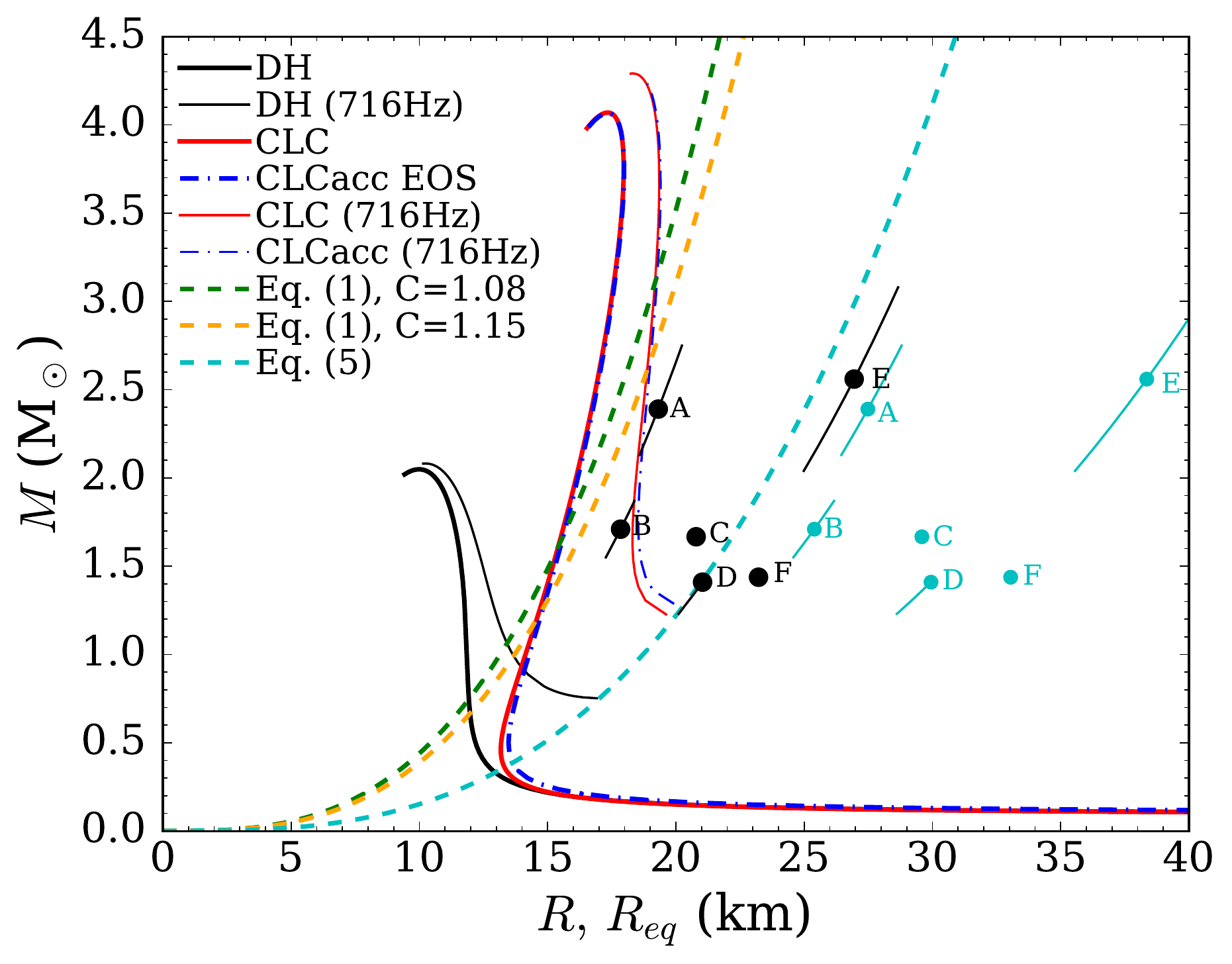}
      \caption{Mass-radius relations for non-rotating stars for the CL EOS (red) with the crust calculated in \cite{DH} (DH) connected at $\rho_u=\rho_0$. For reference we plot the $M-R$ relation obtained for the DH EOS. Dashed lines 
      correspond to Eq.~(\ref{eq:fmax}) for $f=716$ Hz (with two values of the factor $C=1.08$, appropriate for standard hadronic matter, and $C=1.15$, which is used for self-bound stars and is also appropriate for our maximally stiff CL EOS) and Eq.~(\ref{eq:bhz2007}) for $f=716$ Hz (the radius is here the equatorial radius of a rotating star). The crossing point between the theoretical $M(R)$ curves and the dashed curves corresponds to the minimum mass for which the mass-shedding frequency of a star described by the chosen EOS could be $f=716$ Hz. Black points are observational limits from pulsars for which there are simultaneous frequency measurements and mass determinations, with the error bars determined by the uncertainty on mass, obtained by applying Eq.~(\ref{eq:fmax}), whereas cyan points correspond to Eq.~(\ref{eq:bhz2007}). The letters denote the following rapidly-spinning pulsars: A - B1957+20 (622.12 Hz), B - J1023+0038 (592.42 Hz), C - J1903+0327 (465.14 Hz), D - J2043+1711 (420.19 Hz), E - J1311-3430 (390.57 Hz), F - J0337+1715 (365.95 Hz). Theoretical $M(R)$ relations should be located on the left side of the observed pulsar to be consistent with its parameters. The more constraining points correspond to faster pulsars, so that the detection of objects rotating at frequencies above $\sim 1$ kHz would begin to constrain our CL EOS and more generally the physics of dense matter in the stellar interior.}
       \label{FigMR_obs}
   \end{figure}

   \begin{figure}
   \centering
  \includegraphics[width=\hsize]{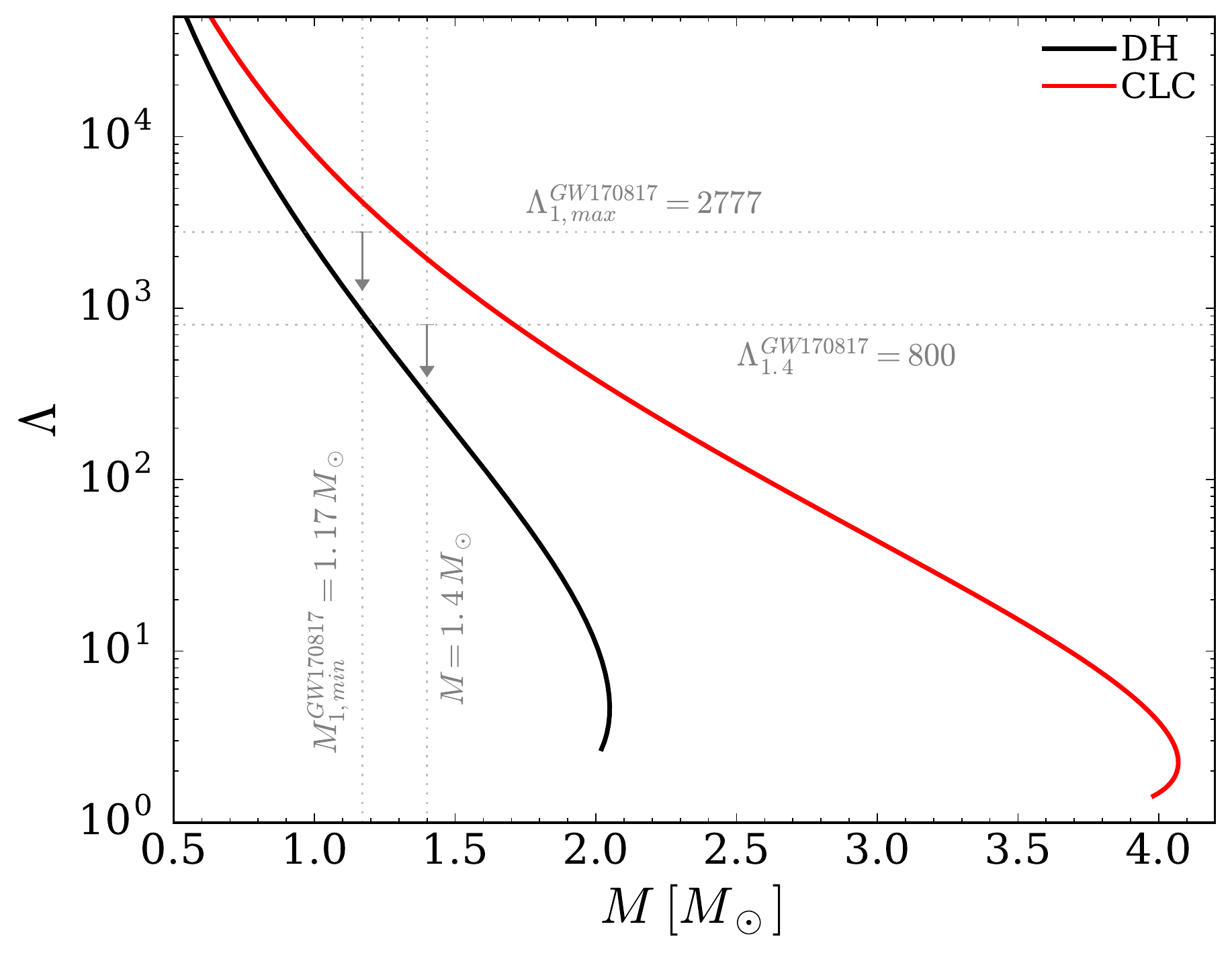}
      \caption{Tidal deformability $\Lambda$ as a function of mass for the CLC EOS (red) with the DH crust connected at $\rho_u=\rho_0$ {(results obtained with the CLCacc EOS are indistinguishable)}, and for the DH EOS (green). The bottom horizontal line corresponds to the limit on the deformability of a $1.4\msun$ obtained from the observations of GW170817, the merger of two NSs: $\Lambda\leq800$. While the DH EOS is compatible with this observational constraint, the  maximally stiff CLC EOS gives much larger values of $\Lambda$ for $M=1.4\msun$. {The top horizontal line corresponds to the tidal deformability in the case in which one of the components of the GW170817 merger was a black hole (or a very compact self-bound star) with vanishing deformability. This limiting $\Lambda_{1,max} = 2777$ was computed from the observational bound on $\tilde{\Lambda}=800$ (Eq.~
\ref{eq:lambdatilde}) for marginal values of the component masses, $M_1 = 1.17\,M_\odot$ (assumed NS, $\Lambda\neq 0$) and $M_2 = 1.6\,M_\odot$ (assumed BH, $\Lambda\equiv 0$) in the low-spin prior estimation case (see Table 1 in \citealt{GWNS} for details).}}
       \label{FigLove}
   \end{figure}

The resulting $M(R)$ plots for both the bare CL EOS NS and the CLC EOS (CL EOS with a realistic crust) NS are presented in Fig.\;\ref{FigMR}. The maximum masses for these EOSs are extremely large as expected for these maximally stiff EOS, much larger than the ones obtained for a realistic EOS like the one from \cite{DH}. The results for the maximum frequency are presented in Fig.\;\ref{FigfM}. The solid lines correspond to calculations obtained using the numerical library {\tt LORENE}, while the dashed ones using the approximation in Eq.\;(\ref{eq:fmax}). The comparison between the results for the CL EOS with a crust confirms that the approximate formula works well up for masses smaller than $0.9\mmax$. 

It is clear that the crust has a strong effect as it significantly increases the radius for a given mass compared to a bare CL star, thus decreasing the maximum frequency $f_k$. The absolute maximum frequency, corresponding the maximum mass for our CLC EOS (i.e. for $\rho_u=\rho_0$), is $f_k^{\mathrm{max}}=1200$ Hz.  This is the lowest value of the maximum frequency for a hadronic star that is still consistent with our understanding of physics below nuclear saturation density and with minimal physical assumptions for the core, namely that the equation of state remain causal.  We note that as the CL EOS is the stiffest possible, any realistic EOS such as the DH EOS will lead to a higher value of $f_k$ at given mass. If continued observations of NS spins confirm the absence of NSs spinning at frequencies higher than $f\approx 700$ Hz, it can be excluded that this corresponds to $f_k^{\mathrm{max}}$, and additional physics is required. 

As an example in Fig. \ref{FigMR_obs} we show $M(R)$ for two equations of state, the realistic DH equation of state and our `minimal' CL equation of state with crust. We also plot the limit given by 
Eq.~(\ref{eq:fmax}) for the currently observed maximum frequency of $f=716$~Hz (dashed lines).  The crossing points between this theoretical curve and the $M(R)$ curves
give us the minimum mass for which 716 Hz could correspond to the maximum frequency. It is about $0.8\msun$ for DH model and  $1.2\msun$ for the CL EOS. The points correspond to pulsars for which there are estimates of both mass and frequency, as given by \citet{Haensel2016} and \citet{Ozel2016}.
The  uncertainty on the mass determination is marked by the fragment of a curve defined by Eq. (\ref{eq:fmax}), for 
the measured frequency of a given object. We see that if NSs rotating at higher frequency were observed, they would begin to provide constraints for models of the high density interior.

{Our conclusions depend very weakly on the assumptions on the crustal EOS we adopt for the NS (whether it is accreted or catalyzed, and which specific model is used). For our analysis the basic parameter of the crust which determines the NS Keplerian frequency is its thickness. The latter depends mainly on the value of the baryon chemical potential $\mu$ at the crust-core interface and at the surface \citep{Zdunik2017}. The detailed description of dense matter in the crust, its composition in particular, influences the microscopic properties such as the heat transport, superfluidity, pinning etc\ldots, but not the thickness in a strong enough way to influence our results. The difference in the radius of a non-rotating NS with an accreted crust and with a catalysed one is $\Delta R(f=0\, {\rm Hz})\sim 100$~m \citep{Zdunik2017} as can be seen in Fig. \ref{FigMR} for the CLCacc model where the accreted crust from \cite{HZ08} is connected to the CL EOS at $\rho_0$. 
Rotation increases the difference between NSs with accreted and catalysed crusts: $\Delta R(f=500\, {\rm Hz})\sim 120$ m and $\Delta R(f=716\, {\rm Hz})\sim 200$ m which amounts, however, to only $\sim 1\%$ of the NS radius, see Fig.~\ref{FigMR_obs}. 
For example for Keplerian configurations with $M=2\,M_\odot$ the frequency for a catalysed crust (corresponding to a more compact NS) is only larger by $\sim 0.6\%$ and the radius smaller by $\sim 0.4\%$ compared to a star with an accreted crust - see Fig. \ref{FigfM} where the CLC and CLCacc curves are almost indistinguishable. Our conclusions on the minimum cutoff frequency for accreting NSs are therefore not influenced by current uncertainties on the crust EOS. }

\section{Conclusions}

In this paper we have examined the problem of whether, given our lack of understanding of high density physics, the observed limit on the rotation rate of NSs of $f\approx 700$ Hz \citep{Patruno2017} can correspond to the maximum rotation frequency of the star, or additional spin-down torques (due e.g. to GW emission or additional spin-down torques in a magnetised accretion disc) must be invoked.

We do this in an EOS independent way, by following the approach of \citet{Koranda1997} and making only minimal physical assumptions, namely that our models remain causal in the high density interior, and that the EOS below nuclear saturation density $\rho_0$ is given by the realistic model of \citet{DH}. This produces a maximally stiff equation of state that will give the lowest possible breakup frequency for a NS. Any softening (as all realistic models will provide) will lead to a higher maximum frequency.

We find that the maximum rotational frequency for a NS cannot be less than $f_{\rm max}\approx 1200$ Hz, and that the observed lack of NSs spinning faster than $\approx 700$ Hz is not consistent with minimal physical assumptions on hadronic physics. Additional mechanisms must be at work to explain this.

The first possibility is the one that is usually considered, i.e. the presence of additional spindown torques acting on the NS, either due to GWs or to interactions between the disc and the magnetic field of the star \citep{Nils1998, Bildsten1998, ABCPHD}. 
Another possibility however, emerges if we observe Fig. \ref{FigfM}. We see that for masses $M>2 M_\odot$ the maximum frequency $f_k\gtrsim 810$ Hz.  The currently observed limit for the spin frequency would thus be roughly consistent with our CL EOS+crust if we are not observing stars with $M\gtrsim 2 M_\odot$. We remind the reader, however, that this is only the case for our CL core EOS, and that realistic EOSs with maximum mass close to $\mmax\approx 2 M_\odot$ are generally much softer and predict much higher values of $f_k$. It would thus be necessary for the EOS to be very stiff (close to the CL EOS) and thus predict a high maximum mass. A possibility is that there are selection effects that prevent us from observing higher mass NSs.
One such possibility might be related to the binary evolution of the LMXB where the recycling process occurs. \citet{TLK12} suggested that the equilibrium spin period that a NS can reach during the recycling process is a steep function of the total mass accreted. The mass transfer process (from the donor star to the surface of the NS) is almost certainly not conservative, since there is evidence for the presence of mass ejection phenomena in LMXBs (e.g., relativistic jets, donor ablation, accretion disk winds). This in turn reduces the maximum amount of mass available for the spin-up, possibly limiting the maximum spin frequency that NSs can reach. However, we note that in this case one would expect a smooth decrease in the number of fast NS, and not the existence of a fast population as observed in accreting systems. Furthermore recent observations by \citealt{Linares} suggest a mass $2.3 M_\odot$ for a recycled NS.

Another intriguing possibility is that the EOS is indeed very stiff, close to the CLC EOS, but there is significant softening at high densities, leading to back bending in the EOS, which leads to a collapse for masses higher than $M\approx 2 M_\odot$, either directly to a black hole, or to a stable branch of hybrid stars \citep{Gerlach68, GK2000, Bejger2017twins}.
Note however that if the system collapses to a more compact configuration, one may expect it to be more rapidly rotating if angular momentum is conserved, and such systems are not observed, although the dynamics of such a collapse are poorly understood. Future work should aim to evaluate the viability of this scenario in more detail.

Furthermore this model would still require the equation of state for hadronic matter to be close to the causal limit, much stiffer than what most models predict and in tension with the limits set by the recent measurements of tidal deformability obtained in GWs from the merger of two NSs, event GW170817 \citep{GWNS}  (see also \citealt{PAS2017} and \citealt{Lattimer18} for a recent analysis). {For this event one has a constrain $\tilde{\Lambda}\le 800$ ($90\%$ credible interval) for a low NS spin prior, where 
\begin{equation} 
  \tilde{\Lambda} = \frac{16}{13}\frac{\left(M_1 + 12M_2\right)M_1^4\Lambda_1 + \left(M_2 + 12M_1\right)M_2^4\Lambda_2}{\left(M_1+M_2\right)^5},
  \label{eq:lambdatilde} 
\end{equation}
can be translated to the deformability of a single NS of $1.4\,M_\odot$, leading to  $\Lambda_{1.4}\leq 800$ \citep{GWNS}. We compare this limit with the DH EOS and the CLC EOS in Fig.~\ref{FigLove}, where one can see that the CLC EOS is incompatible with it. Additionally, one can assume that one of the components in the GW170817 system was a black hole (or a self-bound star) with a vanishing tidal deformability. In that case, if we assume that e.g., $\Lambda_2\equiv 0$, then 
\begin{equation} 
  \Lambda_1 = \frac{13}{16}\frac{\left(M_1+M_2\right)^5}{\left(M_1 + 12M_2\right)m_1^4}\tilde{\Lambda}.
\end{equation} 
To estimate how large $\Lambda_1$ one can adopt adopt $M_1 = 1.17\ M_\odot$ and $M_2 = 1.60\ M_\odot$ (meaning that the NS has mass $M_1$ and the black hole mass $M_2$), and arrive at $\Lambda_1 < 2777$. Even in this extreme case the CLC EOS is only marginally compatible with this bound. 
}

This further strengthens our conclusion that our lower limit on the Keplerian frequency is robust, as more realistic equations of state will always lead to higher Keplerian frequencies. The observed lack of NSs spinning faster than $\approx 700$ Hz cannot be a consequence to the physical breakup frequency of the NS, and additional physics must be at work in these systems to prevent the stars from spinning up to higher rotation rates.

\section*{Acknowledgements}

We acknowledge support from the Polish National Science Centre (NCN) via SONATA BIS 2015/18/E/ST9/00577 (BH) and 2016/22/E/ST9/00037 (MB) and from the European Union's Horizon 2020 research and innovation programme under grant agreement No. 702713 and No. 653477. Partial support comes from PHAROS, COST Action CA16214. AP acknowledges support from an NWO (Netherlands Organization for Scientific Research) Vidi Fellowship. JLZ and MF were supported by the Polish National Science Centre (NCN) grant UMO-2014/13/B/ST9/02621.

\bibliographystyle{aa}

\bibliography{kepler}
\end{document}